\documentclass[prl,twocolumn,showpacs,floatfix,amsmath,amssymb,superscriptaddress]{revtex4}
\usepackage[dvips]{graphicx}
\usepackage{latexsym}
\usepackage{graphicx}
\usepackage{times}
\usepackage{amsmath}
\usepackage{dcolumn}
\usepackage{latexsym,amsmath,amssymb,bm,euscript}
\def\barpsi{\overline{\psi}} 
\def\CsCuCl{$\text{Cs}_2\text{CuCl}_4$} 
\def\CsCuBr{$\text{Cs}_2\text{CuBr}_4$} 
 
\def\DM{\text{Dzyaloshinskii-Moriya}}

\begin{document}

\title{Critical spin liquid at 1/3 magnetization in a spin-1/2
  triangular antiferromagnet} 
\author{Jason Alicea} 
\affiliation{Physics Department, University of California, Santa 
Barbara, CA 93106} 
\author{Matthew P. A. Fisher} 
\affiliation{Kavli Institute for Theoretical Physics, University of 
California, Santa 
Barbara, CA 93106} 
 
\date{\today} 
 
\begin{abstract} 
Although magnetically ordered at low temperatures, 
the spin-1/2 triangular antiferromagnet \CsCuCl~exhibits remarkable 
spin dynamics that strongly suggest 
proximity to a spin liquid phase.  Here we ask whether
a proximate spin liquid may also occur in an applied magnetic field,
leaving a similar imprint on the dynamical spin correlations of this material.
Specifically, we explore a spatially anisotropic Heisenberg spin-1/2 triangular
antiferromagnet at 1/3 magnetization from a dual vortex perspective, 
and indeed find a new ``critical'' spin liquid phase described by
quantum electrodynamics in (2+1)-dimensions with an 
emergent SU(6) symmetry.  A number of nontrivial
predictions follow for the dynamical spin structure factor in this
``algebraic vortex liquid'' phase, which can be tested
via inelastic neutron scattering.  We also discuss how
well-studied ``up-up-down'' magnetization plateaus can be captured
within our approach, and further predict the existence of a stable
gapless solid phase in a weakly ordered up-up-down state.
Finally, we predict several anomalous ``roton'' minima in the
excitation spectrum in the regime of lattice anisotropy where the canted Neel
state appears.

\end{abstract} 
 
\maketitle 
 

The quest for an unambiguous experimental realization of a quantum spin 
liquid remains a central pursuit in condensed matter physics, despite
a long history dating back to Anderson's suggestion that
the nearest-neighbor Heisenberg triangular antiferromagnet may realize
a ``resonating valence bond'' ground state.\cite{RVB}  Quite
generally, the theoretical
search for models realizing such exotic quantum ground states has
focused primarily on frustrated magnets in zero magnetic field.
The central goal of this paper is to 
take a first step at
analyzing the situation when a finite magnetic field is present and 
ask to what extent 
spin liquids may occur in this broader setting.  Naively speaking,
this may seem somewhat misguided, since sufficiently strong magnetic
fields quench quantum fluctuations entirely and lead to a
simple ferromagnetically ordered ground state.  However, at
moderate-strength fields it is conceivable that the presence of
numerous competing phases arising from the geometric frustration may
lead rather to an \emph{enhanced} role of quantum fluctuations by the
field.  

The spin-1/2 triangular antiferromagnet provides a simple and 
experimentally relevant test case for these ideas.  One noteworthy
example is the spatially anisotropic material 
\CsCuCl,\cite{ColdeaHamiltonian,ColdeaShort,ColdeaLong}, which has
garnered much attention as a promising
experimental realization of a spin liquid at zero magnetic field.  
Though \CsCuCl~exhibits long-range spiral order at the lowest
temperatures, inelastic neutron scattering experiments reveal
broad regions of continuum scattering at intermediate energies
throughout the Brillouin zone.\cite{ColdeaLong}  
This anomalous scattering coexists
with well-defined spin-waves in the ordered phase and, significantly,
also persists at temperatures above the Neel temperature where the
spin waves are absent.  The origin of this continuum scattering
presents one of the foremost challenges for understanding the behavior
of this interesting material.  While two groups find 
that nonlinear spin-wave 
theory can account for much of the observed weight in the ordered
phase,\cite{Veillette2, Dalidovich}
spin-liquid physics has been widely invoked as a possible explanation
for the
continuum.\cite{Isakov,Z2anisotropic,SpinonStatistics,ZhouWen,AVLshort,AVLlong,Sheng,SorellaSL}

Low-temperature magnetic order develops in the presence of a 
magnetic field as well, leading to a rich phase diagram.\cite{ColdeaBfield}  
Motivated in part by the observed zero-field phenomenology, we will 
explore the following question here.  Can
spin liquid phases appear at non-zero magnetic field which influence the
dynamics of \CsCuCl~at intermediate energies, just as appears to be the
case in the absence of a field?  Several experimental
features\cite{ColdeaBfield} are 
worth noting that make this scenario plausible.  First, a broad
temperature range characterized by short-range order persists up to
sizable fields of around 6 T.  Second, the
ordering temperature initially \emph{decreases} with the magnetic
field strength,
thereby broadening this short-range order regime.  Finally, there is a
stark contrast in the experimentally determined phase diagrams for
moderate fields applied along the $b$ and $c$ axes in the plane of the
triangular layers, implying a high-sensitivity of ordering to small
perturbations.  
These features together strongly point to the presence of 
many nearly degenerate states and a corresponding enhancement of
quantum fluctuations by intermediate-strength fields.  

\begin{figure}  
  \begin{center}  
    {\resizebox{8cm}{!}{\includegraphics{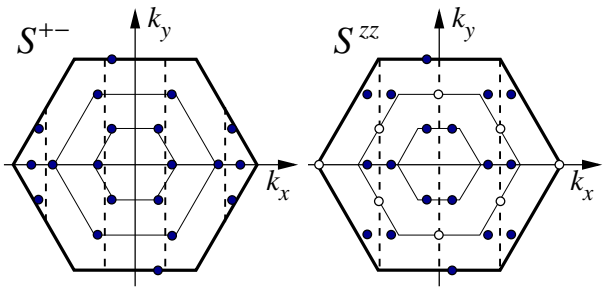}}}  
  \end{center}  
  \caption{Filled circles indicate momenta at which the components of
    the spin structure perpendicular to the field ($S^{+-}$) and along
    the field ($S^{zz}$) exhibit enhanced universal correlations with
    the same power-law decay in the AVL.  In the canted Neel phase,
    anomalous ``roton'' minima in the excitation spectrum are predicted
    at the momenta denoted by open circles.}  
  \label{MonopoleQs}  
\end{figure}

To explore possible field-induced spin liquid phases relevant for
\CsCuCl, we 
study an anisotropic Heisenberg spin-1/2 triangular
antiferromagnet tuned to 1/3 magnetization using a well-studied
duality mapping\cite{duality}.
The present paper significantly extends an earlier study 
of an (easy-plane) spin model
in the absence of a magnetic field.\cite{AVLshort}  At zero-field, it was
argued that reformulating the spin model in terms of 
\emph{fermionized} vortices leads naturally to a spin liquid of the
``critical'' (versus topological) variety.  This 
``algebraic vortex liquid'' (AVL) is a 
promising candidate for explaining the zero-field \CsCuCl~phenomenology.
As we describe below, generalizing to the case of 
1/3 magnetization again 
leads naturally to a ``critical'' AVL phase, which like its zero-field
predecessor supports gapless spin excitations and power-law
spin correlations.  The new AVL characterized here is described 
by quantum electrodynamics in (2+1)-dimensions (QED3) with 
an emergent SU(6) symmetry, which has important
implications for the spin dynamics.  Specifically,
it follows that both the components of the 
dynamical spin structure factor along the field and perpendicular to it 
exhibit enhanced universal correlations with 
\emph{identical} power-laws at various momenta in the Brillouin zone 
(\emph{i.e.}, the filled circles in Fig.\ \ref{MonopoleQs}).  
Such anomalous scattering would be quite interesting to
search for in \CsCuCl~via inelastic neutron scattering.  On a 
technical level, we emphasize two appealing features of this AVL.  
First, the field explicitly breaks the SU(2) spin symmetry down to U(1), so
that our dual description is valid even in the absence of a
\DM~interaction, thereby avoiding subtleties encountered at
zero field.  Second, the QED3 theory describing this AVL is 
``larger-$N$'' than in the zero-field case, and thus even more likely
to exist as a stable phase.  

We also discuss the so-called ``up-up-down'' (UUD) 1/3 magnetization plateaus 
from our dual perspective.  Such UUD states have been
well-studied for the isotropic triangular antiferromagnet 
using spin-wave theory\cite{Chubukov} and exact
diagonalization\cite{Honecker}, and have recently
been observed in the anisotropic material \CsCuBr\cite{CsCuBrPlateau}.  
We show how the 1/3 magnetization plateau can be captured within our approach 
when the sublattice magnetization is near
full polarization, and further predict the presence of a ``critical'' solid
when the sublattice magnetization is weak.  
Signatures of this gapless solid phase may be accessible in exact
diagonalization and/or series expansion studies by examining the
excitation spectrum in an XXZ model with spin-space anisotropy tuned 
near the transition to the UUD plateau.  

Finally, we discuss the range of lattice anisotropy where the
ground state is the canted square-lattice Neel phase.  In this
``frustrated square lattice'' limit, we predict 
several anomalous minima in the excitation spectrum at the
momenta indicated by open circles in Fig.\ \ref{MonopoleQs} which are
due to vortex-antivortex ``roton'' excitations.  These 
excitations can be difficult to capture within spin-wave theory but
appear naturally in our dual framework.  Such features were
also predicted at zero-field, in agreement with earlier series
expansions,\cite{ZhengSeriesExpansion} and would be interesting 
to probe here as well to (perhaps) further substantiate the vortex 
interpretation of those results.

\begin{figure} 
  \begin{center} 
    {\resizebox{6cm}{!}{\includegraphics{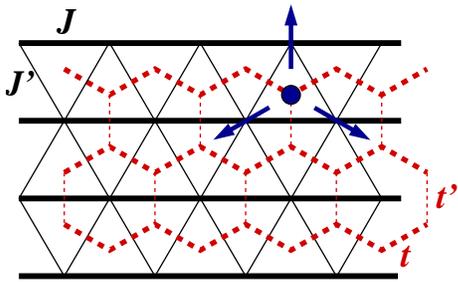}}} 
  \end{center} 
  \caption{(Color online) Triangular lattice and the dual honeycomb 
    on which vortices reside.  Spins shown illustrate a vortex.  } 
  \label{lattice} 
\end{figure}

Turning to the details, the Hamiltonian we consider is
\begin{equation} 
  {\mathcal H} = \sum_{\langle{\bf r r'}\rangle}J_{\bf r r'} {\bf S}_{\bf r}\cdot
  {\bf S}_{\bf r'} + h\sum_{\bf r} S^z_{\bf r}, 
  \label{H} 
\end{equation} 
where the magnetic field $h$ lies in the triangular
planes and the anisotropic exchanges are as shown in Fig.\
\ref{lattice}.  Throughout we assume that the field is tuned so that
the system is at 1/3 magnetization.  Furthermore, we will ignore the
small interplane and \DM~couplings present in \CsCuCl, which is 
appropriate for the field orientations of interest\cite{Starykh}.
Since the Hamiltonian has U(1) spin symmetry, Eq.\ (1) can be mapped 
onto a quantum rotor model and dualized using the standard 
transformation \cite{duality}.  
Readers interested in the details of 
this transformation and the analysis to
follow are referred to Refs.\ \cite{spin1} and \cite{AVLlong}, where
the approach we employ has been well-developed in similar
settings.  Here we will apply this
technique to the new physical situation of a 1/3-magnetized triangular
antiferromagnet, highlighting the nontrivial physics it enables us to
access but dispensing with unnecessary formalism developed elsewhere.  

The quantum rotor mapping is 
implemented by replacing $S^+_{\bf r}\rightarrow e^{i\varphi_{\bf r}}$ and 
$S^z_{\bf r}\rightarrow n_{\bf r}-1/2$, where $n_{\bf r}$ is an 
integer-valued boson number and $\varphi_{\bf r}$ is the conjugate
phase.  While not exact, such a transformation is
expected to be inconsequential for describing the universal physics
that is our focus.  The rotor Hamiltonian can then be expressed as
\begin{eqnarray}
  \mathcal{H}' &=& \sum_{\langle{\bf r r'}\rangle} J_{\bf r
r'}\cos(\varphi_{\bf r}-\varphi_{\bf r'})+U\sum_{\bf r}(n_{\bf r}-1/3)^2
  \nonumber \\
  &+& \sum_{\langle{\bf r r'}\rangle}J_{\bf r r'}(n_{\bf
r}-1/3)(n_{\bf r'}-1/3),
  \label{rotorH}
\end{eqnarray}
where the $U$ term above enforces energetically the constraint of
having either 0 or 1 boson per site as appropriate for modeling a
spin-1/2 system.
In the quantum rotor language, the condition of
1/3 magnetization translates into having on average one boson for every
three sites, which is manifest in the above Hamiltonian.  

The duality mapping applied to Eq.\ (\ref{rotorH}) proceeds
in an identical fashion as in Refs.\ \onlinecite{spin1} and
\onlinecite{AVLlong}; the only difference between these references and
the present system is that the bosons are now at a different mean filling.  
In the dual picture, one equivalently reformulates the rotor model in
terms of quantum mechanical, bosonic vortex degrees of freedom, which are 
topological defects in which the
spins wind around triangular plaquettes as shown in Fig.\
\ref{lattice}.  These vortices are mobile, point-like particles that hop
on the dual honeycomb lattice (see Fig.\ \ref{lattice}) 
in a background of a fluctuating gauge
field $a_{\bf x x'}$ whose flux encodes the boson number (or
equivalently the $S^z$ component of spin, along the field), 
$n^z_{\bf r} \sim (\Delta \times a)_{\bf r}/(2\pi)$.  Thus, the
magnetic field manifests itself as a nontrivial background flux
``felt'' by the vortices, which at 1/3 magnetization is a commensurate
$2\pi/3$ flux per dual hexagon on average.  This background flux is
where the present study departs from the zero-field analysis of Refs.\
\onlinecite{AVLshort,AVLlong}, where the vortices see $\pi$ flux, and
is responsible for the new physics we obtain here.  
The vortices interact via a 
logarithmic repulsion mediated by the gauge field, and importantly are
at half-filling due to the underlying frustration in the original spin
model.  In terms of a
vortex number operator $N_{\bf x}$ and its conjugate phase
$\theta_{\bf x}$, the dual Hamiltonian takes the form 
\begin{eqnarray}
  {\mathcal H}_{\rm dual} &=& -\sum_{\langle{\bf x x'}\rangle} 
  t_{\bf x x'}\cos(\theta_{\bf x}-\theta_{\bf x'}-a_{\bf x x'})
  \nonumber \\
  &+& \sum_{\bf x x'}(N_{\bf x}-1/2)V_{\bf x x'}(N_{\bf
  x'}-1/2)+{\mathcal H}_a.
\end{eqnarray}
The first term allows vortices to hop across nearest-neighbor
honeycomb sites.  The hopping amplitudes are generally anisotropic since
vortices hop more easily across weak spin links; thus we take $t'/t
\sim J'/J$ (see Fig.\ \ref{lattice}).  The second term encodes the 
vortex repulsion, while
the last describes the gauge field dynamics.

The large vortex density poses a significant challenge for
analyzing the dual theory as it stands.  The strong vortex
interactions, however, actually make the problem more tractable, as
they strongly suppress vortex density fluctuations and lead to
an incompressible vortex fluid.  Vortex exchange
statistics is therefore of secondary importance, and one can proceed by
utilizing a formally exact mapping to convert the bosonic vortices
into fermions bound to $2\pi$ flux tubes.  This statistical flux does
not alter the average flux seen by the vortices, since it averages to
$2\pi$ per hexagon (which is equivalent to zero flux).  Furthermore, 
as argued in detail
in Ref.\ \onlinecite{AVLlong} the incompressibility renders the 
flux attachment \emph{irrelevant} for describing low-energy
physics of the vortex fluid.  This statistical transmutation at low
energies from bosonic to fermionic vortices is at the heart of our 
approach.  

Physically, working with fermions is advantageous because the Pauli
principle allows one to first focus on the vortex kinetic energy 
while still maintaining a good interaction energy.
Thus, we initially consider a
mean-field state where we ``smear'' the $2\pi/3$ background flux (due
to the 1/3 magnetization) uniformly across the lattice, though 
we will relax this assumption later.  Fluctuations around this
flux-smeared state can be systematically controlled as 
discussed below.  
Within this mean-field, one can work out the vortex band structure, 
which captures the
important intermediate length-scale physics.  In particular, we find 
that for $t'/t > 2^{1/3}$ the vortices are gapped, while for 
$t'/t < 2^{1/3}$ they
form a ``critical'' state with six gapless Dirac points.  The
former gapped state corresponds to the canted ``square lattice'' Neel
phase expected when $J'$ is dominant, while the latter is the
mean-field description of the AVL which will be our main focus.

We discuss the gapless regime first.  Expanding around the Dirac
points and including gauge fluctuations around the flux-smeared state, 
one obtains a low-energy
effective theory describing six flavors of two-component Dirac
fermions $\psi_{\alpha}$, $\alpha = 1,\ldots,6$, which are minimally
coupled to a U(1) gauge field $a_\mu$ that mediates the vortex
repulsion.  The low-energy effective theory so obtained is identical to QED3,
\begin{eqnarray}
  {\mathcal L}_{\rm QED3} = 
  \barpsi_\alpha (\partial \!\!\! / - i a \!\!\! / ) \psi_\alpha
  + \frac{1}{2 e^2} (\epsilon_{\mu\nu\lambda}\partial_\nu a_\lambda)^2   + {\mathcal L}_{\rm 4f}.
  \label{LQED3} 
\end{eqnarray}
(Inclusion of the statistical $2\pi$ flux tubes merely leads to
interactions that are irrelevant by power-counting in the above, 
as claimed earlier.)  The first term encodes the linearly 
dispersing kinetic energy for our six flavors of vortices.  The second is the
usual Maxwell term, while ${\mathcal L}_{\rm 4f}$ represents symmetry-allowed 
four-fermion interactions.  Of central importance is whether the
``critical'' nature of the mean-field state survives in the full
interacting theory above.  A partial answer to this question can be
obtained by asking whether any fermion mass terms, which would drive some
type of ordering in the original spin model and lead to
a gap in the vortex spectrum, are allowed by symmetry.  
Despite the loss of time-reversal symmetry
by the magnetic field, the answer is ``no''---the remaining symmetries 
of the original spin model are still sufficient to preclude all 
possible mass terms in Eq.\ (\ref{LQED3}).  This is not the full
story, however, since the four-fermion interactions above, if relevant
in the renormalization group sense, could still 
potentially lead to ordering via spontaneous mass generation.  Hence
to proceed we must assess the role of these terms in the theory.

In the limit of a large number $N$ of fermion flavors, all such four-fermion
interactions are indeed known to be irrelevant, so that 
QED3 realizes a nontrivial stable critical phase.  (See QED3
refs.\ in \cite{AVLlong}.)  While the
critical value of $N$ above which this holds is uncertain,
calculations to leading order in $1/N$ suggest that $N = 6$ relevant
here is large enough.  Hence, we proceed with the
assumption that Eq.\ (\ref{LQED3}) indeed describes a stable critical
phase for our vortices, with an emergent SU(6) symmetry due to
the presumed irrelevance of ${\mathcal L}_{\rm 4f}$.  
In terms of the original spin model, this can be qualitatively 
understood as follows.  The presence of
numerous gapless Dirac points implies that there are many competing orders
in the spin model.  With sufficiently many competing orders
(\emph{i.e.}, at large enough ``$N$''), quantum fluctuations can be so
strong as to disorder the system even at zero temperature.  
The resulting critical phase is
precisely the AVL, which respects all symmetries of the original spin
model and supports gapless vortex excitations and, in turn, gapless
spin excitations as we now discuss.  

The key experimental prediction for this phase lies in the behavior of
the dynamical spin structure factor, since this can be directly probed
with inelastic neutron scattering.  We first discuss the spin
correlations of $S^\pm$, transverse to the field.  
Recalling that $S^z+1/2 \sim (\Delta \times
a)/(2\pi)$, since $S^+$ adds $S^z = 1$ it follows that the
corresponding dual operators are ``monopoles'' which add $2\pi$ gauge
flux.  The added flux gives rise to six additional vortex zero-modes, one for
each fermion flavor, and half of these must be filled to produce a
physical state.
Thus there are 20 leading monopole excitations, which can be shown
to carry the momenta ${\bm \Pi}_j$ displayed in the left side of Fig.\
\ref{MonopoleQs}.\cite{AVLlong}  
Such monopoles exhibit nontrivial power-law correlations, each with 
\emph{identical} scaling dimension due to the emergent SU(6) symmetry.  
Consequently, near each ${\bm \Pi}_j$, the
transverse spin structure factor scales as
\begin{eqnarray}
  {\cal S}^{+-} ({\bf k} \!=\! {\bf \Pi}_j + {\bf q},\, \omega) 
  \sim \; 
  \frac{\Theta(\omega^2 - {\bf q}^2)}
     {(\omega^2 - {\bf q}^2)^{1-\eta_{+-}/2}}.
\label{Scrit}
\end{eqnarray}
Using the leading large-$N$ result,\cite{BKW} the anomalous dimension
is $\eta_{+-} \approx 0.54N-1 \approx 2.2$ for each ${\bm \Pi}_j$.

Consider now the correlations of $S^z$, along the field.
Near zero-momentum $S^z$ appears as the dual gauge flux $\Delta \times a$.  
But the vortex band structure
also allows naturally for ``particle-hole'' excitations which
correspond to vortex currents, and such currents generate modulated gauge
flux that contributes to $S^z$ at other wave vectors.  These appear in
the continuum as fermion bilinears, which provide the leading $S^z$ spin 
correlations at the 18 momenta denoted by filled circles in
the right side of Fig.\ \ref{MonopoleQs}.  Near each of these momenta the 
structure factor $S^{zz}$
scales as in Eq.\ (\ref{Scrit}), but with a larger anomalous
dimension $\eta_{zz} \approx 2.3$ estimated from the leading $1/N$
result.\cite{Rantner}     
Note that the momenta denoted by filled circles in Fig.\
\ref{MonopoleQs} shift along $k_x$ as the
lattice anisotropy changes, lining up along the dashed lines in the 1-D
limit and residing symmetrically around the hexagons shown in the
isotropic limit.  We also mention that both $\eta_{+-}$ and
$\eta_{zz}$ are estimated to be larger than 2, in which case a cusp
rather than a divergence occurs in the structure factor.  
Enhanced scattering should nevertheless be observable
near the above momenta, and moreover, in \CsCuCl~such divergences
would be cut off at the lowest energies due to the onset of magnetic order.
 
We now shift our attention to 1/3 magnetization plateaus, 
focusing for simplicity on the isotropic limit $J' = J$.  Returning to
the ``flux-smeared'' mean-field level, we now weakly modulate the gauge flux
away from $2\pi/3$ such that UUD order is assumed at the outset.
Introducing this flux modulation, surprisingly, merely shifts the
locations of the Dirac nodes.  Furthermore, the remaining symmetries
of the spin model are \emph{still} sufficient to protect the
gaplessness of this state against small perturbations.  
That is, the critical nature of the AVL is preserved upon introducing 
weak UUD order, implying the interesting possibility of
having a stable \emph{gapless} solid phase.  

Where, then, is the
gapped magnetization \emph{plateau}?  Such a spin state 
is described by a $\nu = 1$
quantum Hall state for the fermionic vortices.  To see this, note that
the first quantized wave function for the bosonic vortices is 
$\Psi_{\rm bos} = \prod_{i<j}e^{-i \Theta({\bf x}_i-{\bf x}_j)}
\Psi_{\rm ferm}$, where $\Theta({\bf x})$ denotes the angle
formed by the vector ${\bf x}$ with respect to a fixed axis and
$\Psi_{\rm ferm}$ is the fermionic vortex wave function.  
With $\Psi_{\rm ferm}$ the usual $\nu = 1$
wave function, it has been shown\cite{ODLRO} that $\Psi_{\rm bos}$
exhibits (quasi) off-diagonal long-range order.  Hence the bosonic vortices
form a ``superfluid'', which corresponds to an ``insulating'' state
for the original spin model.  Indeed, due to the accompanying 
dual ``Meissner effect'' the gauge field $a_{\bf x x'}$
picks up a Higgs mass, and there are thus no gapless excitations in this
phase.  This is the desired UUD plateau as claimed.  Such a $\nu = 1$ quantum
Hall state can be driven by adding an imaginary second neighbor vortex
hopping with strength $t_2$ larger than a critical value $t_{2c}$,
which decreases as the sublattice magnetization increases towards full polarization.  This can be explicitly demonstrated by computing the Chern numbers
for the occupied bands.\cite{Chern}  Second neighbor
hoppings with $t_2<t_{2c}$ lead to symmetry breaking, 
and are thus precluded in accordance with the above discussion.  

Finally, let us discuss the ``frustrated square lattice'' limit where $J'$
is dominant and the vortices are gapped.  Again, this 
regime corresponds to the expected canted Neel phase.  Consider the
correlations of $S^z$ at momentum ${\bf q}$.  
Aside from spin-waves, the spin structure factor $S^{zz}$ receives
contributions from vortex-antivortex ``roton'' excitations.  In 
the spin language, these correspond to vortex currents generating 
modulated $S^z$ as discussed above in the AVL.  Such excitations are
analogous to Feynman's rotons in He-4, and likewise should appear as 
minima in the structure factor.  The
energy required to create a roton with momentum ${\bf q}$ is
simply given by the minimum energy required to promote a fermionic
vortex in an
occupied band with momentum ${\bf k}$ to an unoccupied band with
momentum ${\bf k-q}$.\cite{AVLlong}  As $J$ increases, enhancing frustration, 
the vortex band gap shrinks leading to a sharp reduction in the
minimum roton energy at the commensurate wave vectors denoted by open
circles in Fig.\ \ref{MonopoleQs}.  
When the gap closes signaling the destruction of the
canted Neel order, the roton energy becomes \emph{gapless} at these
momenta.  (For larger $J$ one enters the AVL phase, and the additional 
momenta in the right side of Fig.\ \ref{MonopoleQs} denoted by 
filled circles then 
branch out from these roton minima.)  The presence of these 
low-energy rotons should lead
to dramatic deviations from linear spin-wave theory.

To conclude, we have provided the first
concrete theoretical proposal for a spin-liquid which may influence
the intermediate-energy dynamics of \CsCuCl~in a magnetic field.
The prospect of observing the spin-liquid physics described
here is an exciting one, and we hope experiments in 
this direction will be pursued.  Our nontrivial 
predictions for the dynamic spin structure factor can be tested
with inelastic neutron scattering by measuring the lower-edge of 
the continuum scattering at the momenta specified in Fig.\
\ref{MonopoleQs}.  Polarized neutrons in
particular would provide a useful probe for the markedly different 
correlations
identified parallel and transverse to the field.  We also
identified a new stable gapless phase with weak UUD order, which 
would be interesting to search for via exact diagonalization
and series expansions.  A renewed look at the excitation spectrum in
the UUD plateau as one adds easy-plane anisotropy to suppress the
solid order may prove fruitful.  Series expansion studies to
search for the predicted rotons in the ``frustrated square lattice''
limit with $J'$ dominant would also be interesting.  More generally,
the new spin-liquid presented here suggests that it may be worthwhile to
widen the search for such exotic phases in other frustrated systems 
by incorporating a finite magnetic field.

\begin{acknowledgments}  
 
It is a pleasure to acknowledge Olexei Motrunich and Leon Balents for   
illuminating discussions. 
This work was supported by the National Science Foundation 
through grants PHY-9907949 (M.\ P.\ A.\ F.) and  
DMR-0529399 (M.\ P.\ A.\ F.\ and J.\ A.).    
  
\end{acknowledgments}


\end{document}